# Survey of LLM Agent Communication with MCP: A Software Design Pattern Centric Review

**Anjana Sarkar**
*Visaze LLC*
anjana@visaze.com

**Soumyendu Sarkar**
*Hewlett Packard Enterprise*
soumyendu.sarkar@hpe.com

**Abstract:** This survey investigates how classical software design patterns can enhance the reliability and scalability of communication in Large Language Model (LLM)-driven agentic AI systems, focusing particularly on the Model Context Protocol (MCP). It examines the foundational architectures of LLM-based agents and their evolution from isolated operation to sophisticated, multi-agent collaboration, addressing key communication hurdles that arise in this transition. The study revisits well-established patterns, including Mediator, Observer, Publish-Subscribe, and Broker, and analyzes their relevance in structuring agent interactions within MCP-compliant frameworks. To clarify these dynamics, the article provides conceptual schematics and formal models that map out communication pathways and optimize data flow. It further explores architectural variations suited to different degrees of agent autonomy and system complexity. Real-world applications in domains such as real-time financial processing and investment banking are discussed, illustrating how these patterns and MCP can meet specific operational demands. The article concludes by outlining open challenges, potential security risks, and promising directions for advancing robust, interoperable, and scalable multi-agent LLM ecosystems.

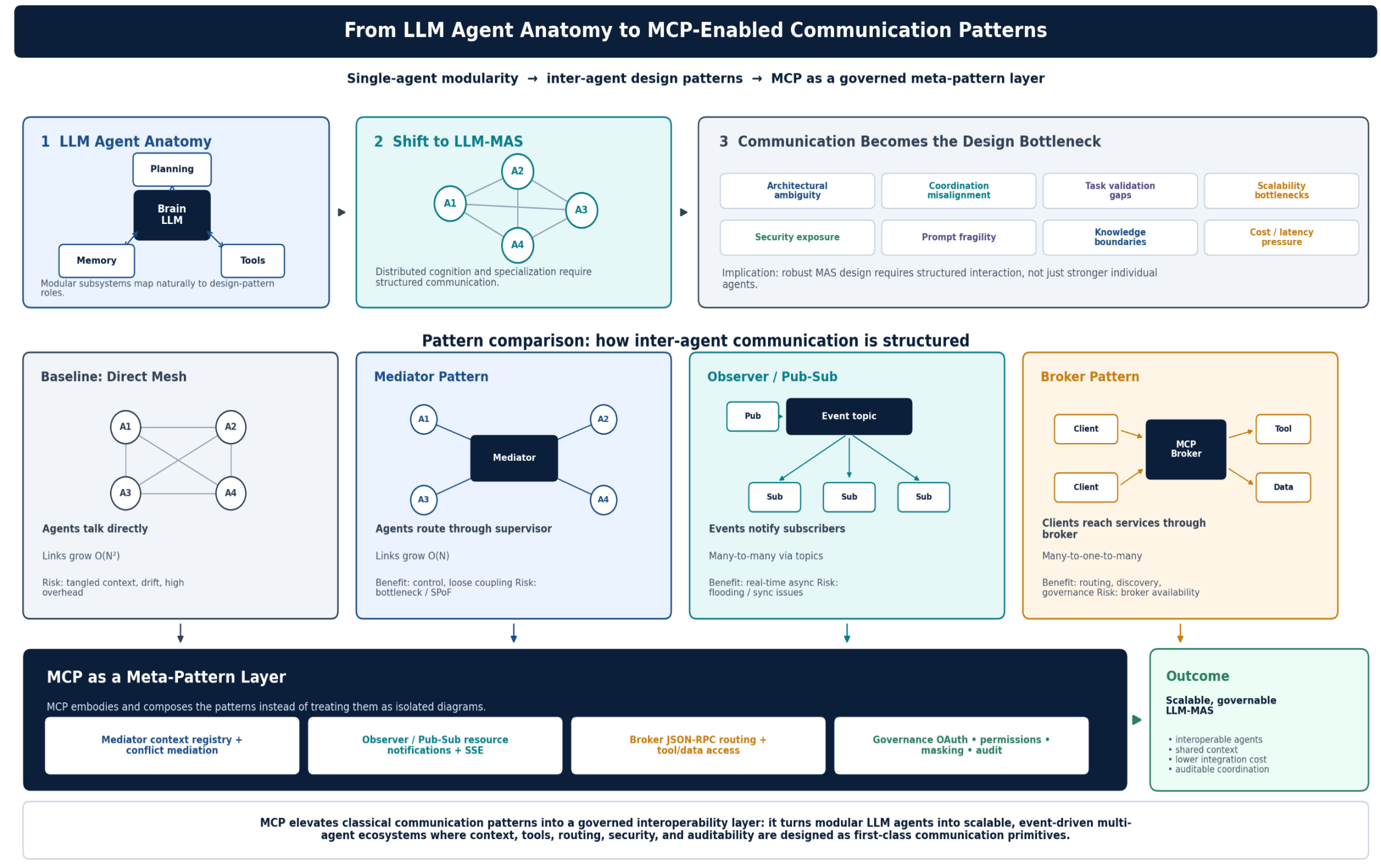

*Figure 1. MCP as a meta-pattern layer for scalable LLM-based multi-agent communication. The diagram traces the architectural progression from a modular single LLM agent, organized around brain, memory, planning, and tool-use components, to multi-agent systems where the dominant design challenge shifts from individual reasoning capability to structured inter-agent communication. It compares direct mesh communication with Mediator, Observer/Publish-Subscribe, and Broker patterns, highlighting their trade-offs in coupling, scalability, coordination, and governance. The central contribution is the positioning of MCP as a unifying meta-pattern layer: rather than serving only as a connector, MCP operationalizes these classical communication patterns through standardized context exchange, resource notifications, tool invocation, routing, security controls, and auditability. This framing shows how MCP can transform fragmented LLM-MAS interactions into scalable, event-driven, and governable agent ecosystems.*

# 1. Introduction

## 1.1. The Emergence of LLM-Based Agentic AI and Multi-Agent Systems

Large Language Models (LLMs) are undergoing a paradigm shift—from functioning as static providers of information, often embedded in conversational agents, to serving as autonomous computational agents capable of decision-making and task execution, often referred to as agentic AI [1]. This shift marks the emergence of agentic AI, wherein LLMs are enhanced with the ability to interact with external systems, store and retrieve information over time, and perform executable actions [2]. These augmented agents are purpose-built to address tasks that require iterative reasoning, planning, memory, and tool use—capabilities that standalone LLMs lack due to constraints like limited context windows, susceptibility to hallucinations, and difficulties in managing complex sequences of actions [3].

As demands grow beyond the scope of a single agent, a new class of systems—Multi-Agent Systems composed of LLM agents (LLM-MAS)—has been introduced. These systems aim to distribute cognitive responsibilities across multiple agents, enabling collaborative problem-solving and specialization [4]. This transition is motivated by the need to scale intelligence through coordinated interactions, especially for real-world tasks [5] that are too complex for individual agents to handle effectively.

Critically, the performance of LLM-MAS is not merely the result of better individual models, but stems from how these agents are architected to communicate, coordinate, and share knowledge [6]. While early LLMs showed strong single-agent performance, they struggled with tasks involving long-term dependencies, contextual continuity, and strategic tool use. Agentic AI addresses these gaps by embedding LLMs within frameworks that support planning, memory, and modular reasoning [4]. However, even these enhancements have limitations when operating in isolation. The transition to multi-agent coordination reflects a recognition that distributed intelligence [7]—emerging from structured, inter-agent communication—is key to tackling high-complexity scenarios. Ultimately, the intelligence exhibited by LLM-MAS arises less from any one agent and more from the system-level design that enables agents to function collectively as a coherent, adaptive unit [8].

## 1.2. The Criticality of Inter-Agent Communication in Complex AI Workflows

Communication between agents is the cornerstone of coordination and shared purpose in multi-agent systems, particularly in those powered by Large Language Models (LLMs). It is through communication that agents align goals, share contextual understanding, and collectively plan actions[5]. However, this very reliance introduces significant challenges. Complexities in inter-agent interaction often contribute more to system-level failures than limitations in the agents themselves. Common difficulties include misaligned objectives, inadequate mechanisms for task validation, limited scalability, exposure to security threats, and the absence of widely accepted architectural standards for robust communication protocols.

In LLM-based multi-agent systems, communication is not just the exchange of information—it is the medium through which collective reasoning emerges. Yet, this strength also becomes a liability: the same communication channels that enable synergy among agents can propagate errors, magnify design weaknesses, and open the door to adversarial exploits such as Agent-in-the-Middle (AiTM) attacks. Thus, communication in LLM-MAS presents a fundamental tension. It is simultaneously the key to emergent intelligence and a critical vulnerability that, if poorly designed, can undermine the entire system. Designing resilient, semantically meaningful communication architectures is therefore not optional—it is central to the success, trustworthiness, and safety of next-generation agentic AI [11].

### 1.3. Introducing the Model Context Protocol (MCP) as an Interoperability Standard

The Model Context Protocol (MCP) [12] [13], introduced by Anthropic in late 2024, is an open interoperability standard aimed at simplifying and unifying the way AI models connect with external tools, systems, and structured data. Often dubbed the "USB-C for AI applications," MCP aspires to be a universal interface layer, reducing the complexity of integration across diverse platforms.

At the heart of MCP is a solution to the long-standing "N x M" integration bottleneck—where each large language model (LLM) required custom code to interface with every distinct data source or tool. This led to duplicated engineering efforts and fragile, difficult-to-maintain architectures. MCP alleviates this by offering a consistent protocol that any AI assistant can use to interact with any compatible service, tool, or dataset, significantly streamlining integration workflows [14].

Built on a client-host-server model using JSON-RPC, MCP enables persistent, state-aware communication sessions. It defines rigorous formats for data ingestion, metadata annotation, platform-agnostic model coordination, and secure bidirectional connectivity. This structured approach not only improves interoperability but also enhances the traceability and manageability of AI-driven systems.

The broader impact of MCP lies in its push toward a modular, composable AI infrastructure. Rather than crafting bespoke connections that quickly devolve into convoluted systems, MCP encourages clean separations between components, allowing tools, models, and data layers to be updated or replaced independently. This modularity greatly reduces engineering overhead, fosters rapid innovation, and provides a foundation for scalable, auditable, and future-proof AI deployments. With clearly defined message schemas and a structured communication lifecycle, MCP also supports critical compliance and monitoring functions—key requirements in enterprise and regulated settings.

### 1.4. Scope and Contributions of this Review: Bridging Design Patterns, LLM Agents, and MCP

This review consolidates recent advancements in large language model (LLM)-driven agentic AI, classical software design methodologies, and the emerging Model Context Protocol (MCP), with the goal of guiding the design of resilient and scalable inter-agent communication frameworks. It examines how time-tested software architecture patterns can be adapted to suit the needs of modern multi-agent systems powered by LLMs, positioning MCP as a core enabler of interoperability and structured coordination.

Through the use of theoretical models and schematic visualizations, the article analyzes communication dynamics, system complexity, and the efficiency of data exchange across agent networks. It also evaluates how these design strategies scale with increasing agent autonomy and system sophistication. Concrete examples are drawn from domains such as real-time financial systems and investment platforms, where robust agent coordination is essential. The review aims to provide developers and system architects with a grounded, actionable framework for building secure, efficient, and maintainable LLM-based multi-agent ecosystems.

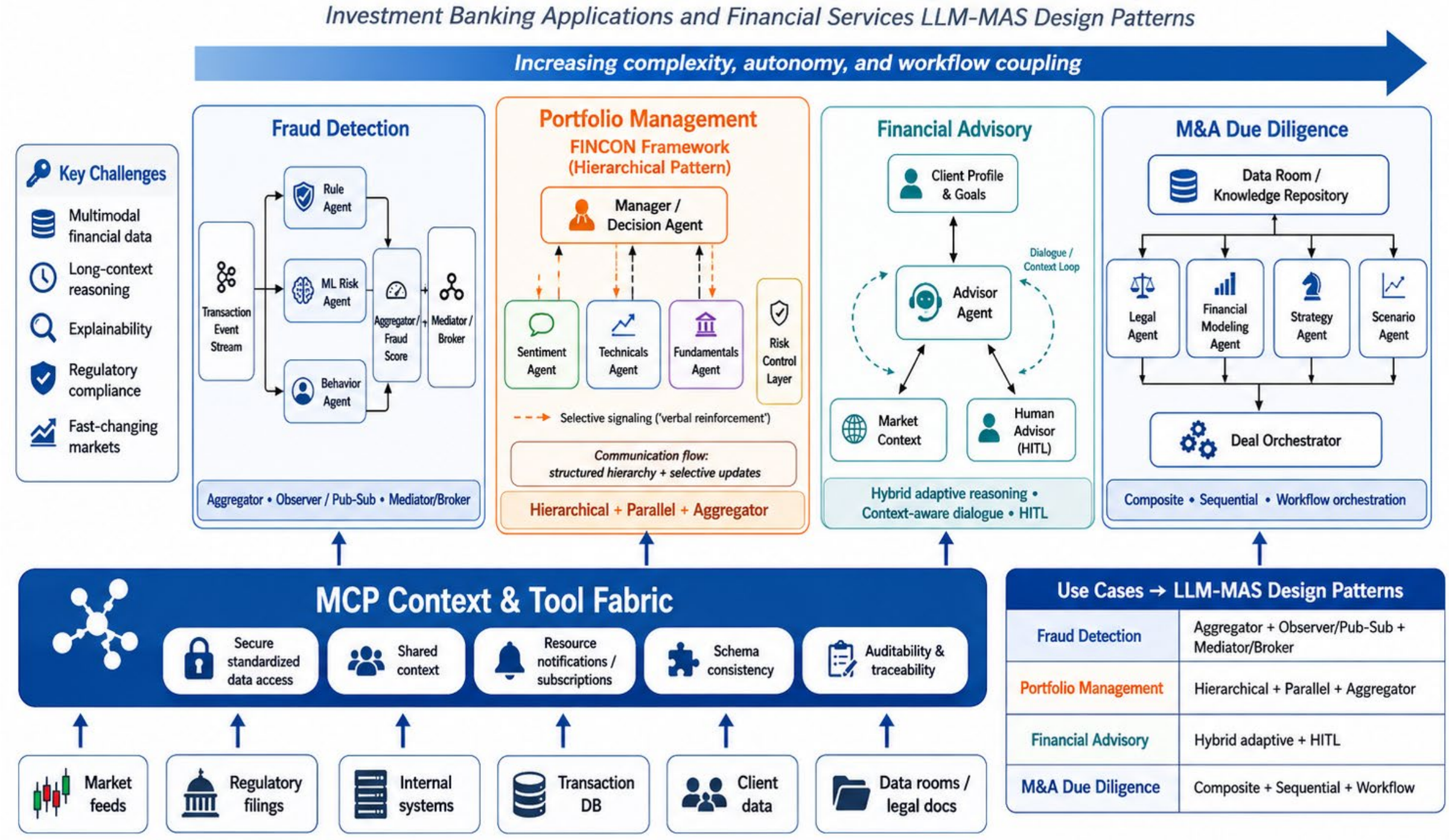

*Figure 2. Architectural adaptation of LLM-based multi-agent systems in financial services. As task complexity increases from the fraud screening to M&A due diligence, interaction patterns evolve from event-driven aggregation to hierarchical coordination and workflow orchestration. MCP acts as the common interoperability layer, grounding agents in secure real-time data, shared context, and auditable communication.*

## 2. Foundations of LLM-Based Agentic Systems

### 2.1. Anatomy of an LLM Agent: Brain, Memory, Tools, and Planning

An LLM-based agent consists of multiple coordinated subsystems that enable it to operate autonomously and interact intelligently with its environment. At the center of the architecture is the large language model itself, which acts as the agent's cognitive core—responsible for reasoning, decision-making, and language comprehension. This central component interprets inputs, forms plans, and produces responses or actions based on its internal logic.

To extend the LLM's capabilities, several auxiliary modules are typically integrated:

- **Memory**: This module plays a key role in sustaining context over time and incorporating insights from past interactions, addressing limitations such as context window size and factual inconsistencies. Memory systems often utilize Retrieval-Augmented Generation (RAG) to supplement the LLM with access to external sources of dynamic or long-term information [15].

- **Planning**: Responsible for breaking down complex goals into actionable steps, the planning module enables the agent to reason through multi-stage tasks. Methods like Chain-of-Thought (CoT) reasoning are employed to promote transparency in intermediate steps and support iterative refinement of plans.

- **Tool Use**: This module enables the agent to interface with external systems and perform targeted operations—such as querying databases, invoking APIs, executing code, or conducting searches. Effective use of tools depends on structured, interpretable definitions that clearly specify each tool's functionality [16] [17].

The modular composition of these components makes LLM agents highly compatible with classical software design patterns. The separation of roles into distinct modules—such as reasoning, memory, planning, and tool usage—naturally aligns with established architectural and behavioral templates. For instance, the tool interface is well-suited to structural patterns like Adapter or Facade, which abstract complexity and promote standardization across diverse external services. The memory subsystem can adopt design principles for data persistence and access control, ensuring efficient and consistent storage and retrieval. Meanwhile, the central LLM or "brain" can leverage behavioral patterns like Mediator or Strategy to coordinate interactions among modules and manage dynamic decision processes.

This intrinsic modularity not only enhances the clarity and maintainability of individual agents but also lays a strong foundation for building scalable, interoperable systems of multiple agents. The ability to clearly delineate responsibilities among components is essential for designing robust internal behaviors and facilitating structured collaboration in broader multi-agent frameworks.

## 2.2. From Single-Agent Autonomy to Multi-Agent Collaboration

Although single-agent systems powered by large language models (LLMs) can perform complex reasoning and execute tasks independently, they often struggle with problems that require distributed cognition or large-scale coordination. These agents are inherently limited by their sequential processing, restricted memory capacity, and the finite bandwidth of a single decision-making entity.

To address these constraints, Multi-Agent Systems (MAS) have emerged as a paradigm that enables multiple intelligent agents to collaborate. By distributing responsibilities and enabling inter-agent communication, MAS architectures facilitate scalable task execution, improve system resilience, and allow for dynamic adaptation in real-time environments. Through coordinated behavior, these systems can demonstrate emergent intelligence—where collective performance exceeds the capabilities of individual agents acting alone. Notably, LLM-based MAS (LLM-MAS) benefit from natural language communication channels, hierarchical task delegation, and integration with domain-specific tools and knowledge sources, often without needing hand-coded rules.

However, transitioning from isolated agents to collaborative systems shifts the primary design challenges. Instead of being constrained by the cognitive limits of a single model, the system's reliability now hinges on the quality and structure of inter-agent interactions. While multi-agent setups offer a promising solution to the shortcomings of single LLMs—such as hallucination, short context retention, and planning bottlenecks—they introduce their own complexities. Empirical studies have shown that the anticipated performance improvements are not always realized, often due to miscommunications, coordination overhead, and agent misalignment. This underscores the importance of reorienting the design focus: success in multi-agent systems depends not only on building capable agents but, critically, on engineering robust, coherent, and efficient communication and coordination mechanisms among them [18].

## 2.3. Inherent Challenges in Multi-Agent LLM Communication

While multi-agent systems powered by large language models (LLMs) offer compelling advantages in collaborative reasoning and distributed task execution, they also introduce a complex set of communication and coordination challenges that must be systematically addressed to ensure reliability and scalability.

- **Architectural Ambiguity**: One of the foundational issues is the absence of standardized frameworks for designing robust LLM-based multi-agent systems (LLM-MAS). This often leads to improvised architectures that lack consistency and resilience [19].
- **Coordination and Misalignment**: Achieving effective collaboration among agents is non-trivial. Agents must be able to engage in joint reasoning, maintain shared context, and align their goals—tasks that are often hindered by incoherent or unstructured communication [20].
- **Task Completion and Validation**: Determining when a task is complete and verifying the accuracy or success of a multi-agent process is inherently difficult in distributed systems, where no single agent has a complete view of the task state.
- **Scalability Bottlenecks**: As more agents are added, communication overhead rises sharply. This leads to increased latency, bandwidth saturation, and higher computational resource demands, reducing system responsiveness.
- **Security Risks**: The decentralized nature of multi-agent communication increases exposure to threats. Vulnerabilities such as Agent-in-the-Middle (AiTM) attacks, data leakage, and injection of malicious prompts pose significant risks to privacy and integrity.
- **Prompt Fragility**: The behavior of LLM agents remains highly sensitive to how prompts are phrased. Small changes in input can lead to drastically different and sometimes unreliable outputs, undermining the system's predictability.
- **Knowledge Management and Hallucination**: Defining and enforcing knowledge boundaries within multi-agent environments is challenging. Without careful control, agents may generate biased, inaccurate, or fabricated information, compromising the credibility of the entire system [21] [22].

Designing LLM-MAS is often likened to managing a human organization, complete with role specialization, hierarchical planning, and collaborative problem-solving. However, this analogy also underscores the system’s inherent complexity. Just as real-world organizations can falter due to structural flaws—like miscommunication, departmental silos, or strategic misalignment—LLM-MAS inherit similar vulnerabilities. The challenges of governance, coordination, and failure modes in such systems are not purely technical but deeply systemic. As a result, building effective LLM-MAS may require integrating insights from organizational science and human systems engineering, emphasizing the importance of structured collaboration, trust frameworks, and robust communication protocols alongside traditional AI techniques.

## 2.4. Comparison of Key LLM Agent Frameworks and their Communication Paradigms

**AutoGen** employs a message-passing paradigm, primarily using broadcast or publish-subscribe mechanisms to facilitate communication among agents. It supports multi-agent dialogues, integrates Large Language Models (LLMs), human inputs, and external tools, and allows for highly customizable interactions. This flexibility makes AutoGen well-suited for building diverse, complex workflows. However, it can be difficult to manage context and ensure consistent alignment between agents, especially as complexity grows.

**LangChain** (and its graph-based extension **LangGraph**) follows a node-and-edge orchestration model where agents and workflows are structured as directed graphs. These frameworks offer capabilities such as memory management (short-term and long-term), human-in-the-loop interaction, and detailed observability. They are particularly strong in enabling both deterministic workflows and dynamic agent orchestration, which makes them highly suitable for complex tasks. Nevertheless, LangChain and LangGraph may complicate the control over LLM context and introduce debugging challenges in more intricate graph configurations.

**CrewAI** is built around a workflow model inspired by human teaming. It emphasizes role-based agents with clearly defined responsibilities, facilitating intuitive collaboration and coordination that mimics real-world teams. This design is advantageous for structured, collaborative tasks that benefit from role clarity. However, it can be less effective in scenarios requiring fluid, adaptive responses outside the bounds of predefined roles, potentially limiting its applicability in highly dynamic environments.

**MetaGPT** leverages principles from software engineering to structure agent workflows. It is particularly well-suited for tasks that require ordered, rule-based collaboration, such as software development. The framework excels at coordinating multiple agents toward a common, structured goal and ensures clarity in output. On the downside, MetaGPT can face limitations in scalability, is vulnerable to bottlenecks or failure points due to centralized coordination, and may not adapt easily to unstructured or evolving task environments.

**Google's Agent Garden**, alongside its **Agent2Agent (A2A) Protocol**, adopts a JSON-based lifecycle model that facilitates peer-to-peer task outsourcing. It provides a centralized hub of pre-built agents and a communication framework that promotes interoperability across diverse technology stacks. This design simplifies enterprise-level integration and supports collaborative task sharing between heterogeneous agents. However, the pursuit of cross-vendor standardization presents non-trivial challenges, and the coordination infrastructure may introduce communication overhead as complexity scales.

While multi-agent systems powered by large language models (LLMs) offer compelling advantages in collaborative reasoning and distributed task execution, they also introduce a complex set of

## 3. Software Design Patterns for Inter-Agent Communication

### 3.1. Re-evaluating Classical Design Patterns for LLM-MAS

Software design patterns [23] are established, reusable solutions that address common challenges in software engineering. They support the creation of systems that are modular, scalable, maintainable, and reusable. Traditionally classified into creational, structural, and behavioral categories, each pattern type focuses on solving a specific design concern. Although these patterns have played a central role in conventional software development, their application within Multi-Agent Systems (MAS) has been relatively limited. This limited adoption is partly due to the absence of standardized documentation practices and a lack of clarity in how different patterns interrelate or should be composed in MAS contexts.

The rise of AI—particularly large language models (LLMs)—is redefining the role of design patterns in intelligent systems. Unlike traditional implementations, which tend to be rigid and static, LLM-based agents offer the ability to reason, reflect, and adapt in real time. This opens the door to a new class of "dynamic patterns," where classical design templates are no longer fixed but can evolve in response to system performance or environmental feedback.

One of the long-standing critiques of design patterns has been their inflexibility in dynamic or fast-changing settings. LLM agents, by contrast, are inherently adaptable and capable of on-the-fly decision-making, enabling patterns to become context-aware and self-adjusting. When integrated into system architectures, these AI agents can refine or reconfigure pattern implementations in real time—modifying behavior, adjusting workflows, or even shifting structural configurations based on observed metrics or predicted needs. This creates a feedback loop in which AI and architecture co-evolve: LLM agents bring responsiveness and learning capacity, while design patterns contribute structure and reliability.

The result is a symbiotic design paradigm—dynamic patterns—where architectural strategies are no longer predefined blueprints but evolving frameworks, shaped and informed by the AI agents they support. This shift holds promise for creating intelligent systems that are not only robust and maintainable but also responsive and continuously improving.

## 3.2. Key Communication Patterns: Mediator, Observer, Publish-Subscribe, Broker

Behavioral design patterns are particularly crucial for defining effective communication and delegating responsibilities within LLM-MAS, ensuring flexibility and scalability without introducing tight coupling between components.

**Mediator Pattern**

- **Intent:** The Mediator pattern aims to reduce chaotic dependencies among objects by centralizing their interactions. Instead of direct communication, objects collaborate indirectly by calling a special mediator object that redirects calls to appropriate components.
- **Application in LLM-MAS:** In multi-agent LLM systems, a mediator agent, often a supervisor LLM, can centralize communication, preventing direct, potentially chaotic interactions among numerous specialized agents. This approach promotes loose coupling, making individual agents easier to modify, extend, or reuse in different contexts.
- **MCP Alignment:** The Model Context Protocol (MCP) can function as a central registry for context versioning and can mediate conflicting actions among agents. This aligns with the Mediator pattern's principles of centralizing interactions and resolving discrepancies. The MCP broker pattern is specifically designed as a flexible, intelligent middleware that facilitates communication between diverse system components [24].

**Observer Pattern / Publish-Subscribe (Pub/Sub) Pattern**

- **Intent:** The Observer pattern defines a one-to-many dependency between objects, where a subject (publisher) automatically notifies all its registered dependents (observers/subscribers) of any state changes. In the Pub/Sub model, publishers are decoupled from subscribers through event channels or a message broker [25].

- **Application in LLM-MAS:** These patterns are ideal for building event-driven architectures and enabling real-time updates within LLM-MAS. Agents can subscribe to specific topics, such as financial news feeds or market data changes, and receive notifications of relevant events without needing to constantly poll for updates or possess explicit knowledge of the message senders. This asynchronous communication mechanism is critical for achieving scalability and responsiveness in dynamic multi-agent environments.
- **MCP Alignment:** MCP [26][27]inherently supports publish-subscribe mechanisms through its resource change notifications and Streamable HTTP implementation, which includes Server-Sent Events (SSE). This functionality allows agents to subscribe to changes in shared resources, enabling the development of sophisticated inter-agent workflows with complex dependencies.

**Broker Pattern**

- **Intent:** The Broker pattern is an architectural pattern that uses an intermediary component, the "broker," to facilitate communication between decoupled components, typically servers (publishers) and clients (subscribers), via remote procedure calls. The broker maintains routing and filter tables and can provide additional functionalities such as Quality of Service (QoS) guarantees or security enforcement [28].
- **Application in LLM-MAS:** This pattern provides a centralized intermediary for asynchronous communication and coordination, significantly promoting loose coupling, scalability, and resilience within distributed systems. MCP servers themselves function as brokers, mediating interactions between LLM clients and various external data sources or tools.
- **Distinction from Pub/Sub:** While similar, the Broker architectural pattern is typically represented by a "Many to One to Many" diagram, indicating a centralized intermediary. In contrast, the Publish-Subscribe architectural pattern is often depicted as a "Many to Many" relationship, where messaging functionalities are often hidden as a cross-cutting concern. MCP itself can be considered an implementation of the Broker pattern.

These behavioral patterns—Mediator, Observer/Pub-Sub, and Broker—form the backbone of structured interaction in LLM-based multi-agent environments. What distinguishes MCP is not just its compatibility with these patterns, but its embodiment of them. MCP functions simultaneously as a coordination center (Mediator), an event notification platform (Observer/Pub-Sub), and a communication intermediary (Broker). This multifaceted role elevates MCP beyond a mere protocol—it becomes a unifying architectural layer that enables consistent, context-aware communication across agents and tools.

This integration results in a layered communication model where MCP operates as a meta-pattern, establishing a foundational transport and protocol layer upon which more advanced interaction strategies—such as negotiation, dynamic task delegation, or collaborative planning—can be reliably built. This hierarchical approach provides LLM-MAS with a scalable and extensible framework for achieving robust, interoperable, and intelligent behavior at both the agent and system levels.

## 3.3. Key Communication Patterns: Mediator, Observer, Publish-Subscribe, Broker

**Mediator Pattern**

The Mediator pattern focuses on centralizing communication between components to reduce direct dependencies and tangled interactions. In LLM-MAS environments, this pattern is commonly realized through a supervisor agent that manages interactions among various specialized agents. This orchestration prevents agents from engaging in ad hoc communication, which can quickly become chaotic and difficult to manage at scale. MCP aligns with this pattern by serving as a centralized context registry and by mediating conflicting actions among agents. Its built-in broker pattern also enables structured communication between heterogeneous components. The benefits of using the Mediator pattern include reduced coupling, improved agent modularity and reusability, and streamlined control logic. However, if not carefully designed, the mediator can become a bottleneck or a single point of failure, especially under high loads or in complex workflows.

**Observer / Publish-Subscribe Pattern**
The Observer pattern, and its event-driven variant Publish-Subscribe (Pub/Sub), defines a one-to-many relationship where a subject automatically notifies observers of state changes. In the context of LLM-MAS, agents often subscribe to streams of events—such as task progress, environmental updates, or financial data—allowing them to receive updates in real time without constant polling or tight coupling to event sources. MCP enables this interaction model through its support for resource change notifications and Streamable HTTP using Server-Sent Events (SSE), making it possible to implement reactive, dynamic context updates. This pattern offers several advantages, including decoupled system components, improved scalability, and the foundation for event-driven architectures with real-time responsiveness. Still, it introduces potential challenges such as message flooding, synchronization difficulties, and memory leaks related to "lapsed listeners" that fail to unsubscribe properly.

**Broker Pattern**
The Broker pattern introduces an intermediary layer to manage communication between clients and servers, enabling loosely coupled interactions while abstracting service discovery and request routing. Within LLM-MAS, MCP servers function as brokers by handling interactions between language model clients and external data tools or APIs. MCP's client-host-server design naturally supports the broker pattern by standardizing tool invocation, managing context access, and routing requests across system components. The main strengths of this pattern are enhanced system scalability, resilience, and modularity, as well as a centralized communication interface that simplifies integration. Nevertheless, similar to the Mediator, the broker can become a point of failure if not distributed or redundantly implemented. Additionally, ensuring consistent and accurate data transmission through the broker remains a technical challenge, particularly in high-throughput environments.

## 3.4. Formalizing Communication Patterns in LLM-MAS: A Mathematical Perspective

Formalizing inter-agent communication within LLM-MAS provides a quantitative framework for understanding and optimizing system behavior. This involves applying concepts from graph theory and information theory to model communication overhead, information flow, and associated costs.

**Communication Overhead (Graph Theory)**

The complexity of communication links varies significantly with the chosen architectural pattern. Let N represent the total number of agents in a multi-agent system.

- **Fully Decentralized Communication (e.g., Flat Architecture, Network):** In a system where every agent can communicate directly with every other agent , the number of direct communication links, $L_{direct}$, grows quadratically with the number of agents.

$$L_{direct}=O(N^2)\approx N(N-1)/2$$

  This implies that as the number of agents increases, the complexity of managing direct connections and the potential for communication bottlenecks rise rapidly.

- **Centralized Communication (e.g., Mediator, Broker, Supervisor):** In contrast, when all communication is routed through a central entity or intermediary , the number of direct links to this central entity, $L_{centralized}$, grows linearly with the number of agents.

$$L_{centralized}=O(N)\approx N$$

  This mathematical relationship highlights the inherent scalability advantage of centralized patterns in managing communication complexity, as they significantly reduce the number of direct inter-agent connections required.

**Information Entropy in Message Passing (Information Theory)**

The information content and efficiency of messages can be quantified using principles from information theory, as inspired by Shannon's work.

- Let M be the discrete message space. The entropy $H(M)$ quantifies the average uncertainty or information content of messages exchanged within the system.
- The mutual information $I(A;B)$ between two agents, A and B, measures the amount of information about agent A's state that agent B gains through their communication. It is defined as:

$$I(A;B)=H(A)-H(A|B)$$

  where $H(A)$ is the entropy of agent A's state, and $H(A|B)$ is the conditional entropy of A's state given B's observation of the message.

- **Mathematical Implication:** Effective communication patterns aim to maximize mutual information while minimizing redundant or irrelevant messages. Design patterns like Mediator or Broker, by intelligently filtering and routing only the most relevant information, can optimize this by reducing $H(A|B)$ for the receiving agent while simultaneously minimizing the overall volume of unnecessary message traffic.

**Cost of Communication**

The practical deployment of LLM-MAS necessitates consideration of the computational and financial costs associated with inter-agent communication.

- **Mathematical Implication:** Efficient communication patterns, such as selective propagation in Observer/Publish-Subscribe models or the use of summarized messages in Mediator patterns, directly reduce the number of tokens exchanged ($tokens_{i\to j}$). This, in turn, leads to a reduction in $C_{comm}$, which is a critical factor for ensuring the practical and cost-effective deployment of LLM-MAS.

**Strategic Implications**
Quantitative modeling of communication patterns offers more than just theoretical insights—it grounds system design in measurable parameters. The stark difference between O(N2)O(N^2)O(N2) and O(N)O(N)O(N) connection complexities underscores why centralized designs are often favored in large-scale systems. Similarly, applying entropy and cost models helps engineers refine message content and frequency to minimize overhead and latency. This structured, analytical approach elevates architectural decisions from heuristic practices to evidence-based design, ensuring that scalability, efficiency, and affordability are systematically addressed in LLM-MAS development..

## 4. The Model Context Protocol (MCP) as an Interoperability Layer

### 4.1. MCP Architecture: Client-Host-Server Model and JSON-RPC Foundation

The Model Context Protocol (MCP) is built upon a robust client-host-server architecture, designed to standardize communication between AI applications (which function as hosts or clients) and various external resources (acting as servers). This architecture aims to provide a unified and secure interface for AI models to interact with the broader digital environment.

- **Host:** The host is the primary LLM application, such as Claude Desktop or an Integrated Development Environment (IDE), that initiates connections and directly interacts with users. It plays a crucial role in managing security policies, user authorization, and consent requirements, ensuring that AI actions align with user permissions and organizational guidelines.
- **Client:** Embedded within the host application, the client is a lightweight protocol component that maintains a one-to-one connection with a specific MCP server. Its responsibilities include handling capability negotiation with servers and orchestrating messages between the host's LLM and the external resource.
- **Server:** MCP servers are independent processes that expose specific capabilities, such as tools, data access, or prompts, in a standardized manner over the MCP. These servers can operate locally or remotely and act as wrappers around various external systems like APIs, databases, or file systems. Examples of existing MCP servers include those for GitHub, Postgres, Tavily, and Chargebee, among many others.

The foundational communication mechanism for MCP is **JSON-RPC 2.0**. This standard defines clear message types—requests (expecting a response), results (successful responses), errors (indicating failure), and notifications (one-way messages)—and a structured connection lifecycle that includes initialization, message exchange, termination, and error handling. For the **transport layer**, MCP supports both Stdio (for efficient local processes) and HTTP with Server-Sent Events (SSE) (for networked services and remote integrations).

The client-host-server framework underlying MCP, implemented via JSON-RPC, offers a disciplined and transparent interface that helps demystify the often opaque behavior of large language models (LLMs). One of the persistent issues with LLMs is their unpredictability, including their tendency to generate misleading or inconsistent outputs. MCP addresses this by clearly delineating the responsibilities of the client, host, and server, and grounding all communication in a structured protocol. JSON-RPC enforces standardized message types—such as requests, responses, errors, and notifications—and prescribes a defined interaction lifecycle from connection initiation to closure.

As a result, exchanges between the LLM (via the client) and external systems (via servers) are no longer informal or solely dependent on natural language prompts. Instead, these interactions are handled through a formalized protocol that enables greater control and clarity. This structured design greatly enhances the ability to trace, troubleshoot, and validate agent behaviors, effectively mitigating the "black box" challenge. The resulting boost in system transparency and dependability is a key enabler for the adoption of LLM-driven multi-agent systems in high-stakes, enterprise-level deployments.

### 4.2. MCP's Role in Standardized Context Exchange and Tool Invocation

The Model Context Protocol (MCP) serves as a foundational mechanism for standardizing how critical contextual elements—such as tools, datasets, and inference configurations—are delivered to and utilized by large language models (LLMs). Acting as a universal integration layer, MCP streamlines the complexity of connecting diverse components within AI ecosystems.

A key problem MCP addresses is the "N × M" integration challenge, where each combination of model and data source traditionally required bespoke code. MCP resolves this by offering a consistent interface for registering, discovering, and executing tools, thereby eliminating the need for handcrafted connectors across different AI systems [29].

MCP introduces several core capabilities that empower flexible and intelligent multi-agent interactions:

- **Context Sharing**: Through its resource capability, MCP allows agents to exchange data such as files, internal state, or memory. It also supports change notifications on shared resources, enabling agents to build reactive workflows that adjust dynamically to evolving conditions.
- **Tool Invocation**: LLMs can access and invoke external functions—termed "tools"—exposed by MCP servers. These tools may perform operations like calling APIs, running database queries, or executing modular agent functions [30].
- **Sampling**: This capability allows agents to share prompts and, in some cases, even delegate tasks between different LLMs. It creates opportunities for collaborative reasoning and peer-based assistance within an LLM-MAS environment.

Together, MCP's context-sharing and sampling functionalities allow for the emergence of a distributed, real-time knowledge graph that surpasses the limitations of static Retrieval-Augmented Generation (RAG). Traditional RAG approaches supplement a single LLM with pre-indexed information retrieved on demand, but this model remains largely passive and centrally constrained. In contrast, MCP enables agents to continuously publish and subscribe to evolving resources—be it files, state information, or agent memory—creating a shared, adaptive knowledge space. Moreover, sampling extends this capability beyond data, allowing agents to share their reasoning capacity and act as active collaborators. This dynamic architecture transforms the RAG paradigm into a living, distributed framework where context is not just retrieved but jointly managed and evolved—supporting more sophisticated, decentralized multi-agent reasoning systems [31].

### 4.3. MCP as a Facilitator for Inter-Agent Communication Patterns

MCP's Streamable HTTP interface equips multi-agent systems with a versatile set of communication modes. It supports a range of interaction models—from stateless request-response exchanges to persistent, stateful sessions using unique identifiers, as well as real-time streaming through Server-Sent Events (SSE). This flexibility makes MCP a powerful foundation for implementing common software design patterns within LLM-based multi-agent systems (LLM-MAS).

- **Mediator Pattern**: MCP servers can serve as centralized coordinators that handle the routing of messages between LLM clients and various external services or tools. By abstracting the direct connections between agents and external systems, this aligns naturally with the Mediator pattern's goal of reducing tight coupling between components [32].
- **Observer / Publish-Subscribe Pattern**: MCP enables agents to subscribe to notifications for updates on shared resources. This capability mirrors the core idea of the Observer and Pub/Sub patterns, where agents can react automatically to changes in system state, enabling responsive and event-driven collaboration.
- **Broker Pattern**: The MCP architecture itself functions as an intermediary layer that separates agents from the specifics of the underlying tools and data sources. By acting as a centralized broker, MCP simplifies integration and promotes modularity.

These foundational communication mechanisms provided by MCP form the concrete infrastructure that supports the implementation of higher-level coordination strategies described by software design patterns. While patterns like Mediator or Observer define the abstract structure of agent interactions, MCP provides the operational infrastructure that brings these patterns to life in practice. Its use of JSON-RPC messaging, combined with Streamable HTTP and standardized resource and tool capabilities, delivers the essential mechanisms for scalable, real-time collaboration.

For example, implementing the Observer pattern requires a reliable way to notify agents about changes—something MCP enables directly through SSE-based resource updates. Likewise, the Mediator pattern needs a centralized point for managing interactions; MCP servers fulfill this role by standardizing access to tools and data. Thus, MCP is not just a protocol for message exchange—it acts as a foundational framework that makes the execution of sophisticated communication patterns in LLM-MAS both practical and efficient.

### 4.4. Comparative Analysis of Agent Interoperability Protocols (MCP, A2A, ACP, ANP)

In addition to MCP, a range of emerging protocols is shaping the future of agent interoperability, each designed to meet specific requirements across different deployment scenarios [33].

- **Model Context Protocol (MCP)**: MCP is designed to standardize how structured context—such as tools, datasets, and sampling instructions—is delivered to large language models. It operates through a JSON-RPC-based client-server interface and supports Streamable HTTP and Server-Sent Events (SSE). Key capabilities include resource (for sharing files, state, or memory across agents) and sampling (for prompt/model sharing between agents), enabling dynamic and collaborative workflows. Its communication model follows a many-to-one-to-many pattern, making it particularly effective for tool invocation, enterprise data integration, and the creation of deeply stateful agents. Often referred to as the "USB-C for AI," MCP provides foundational support for higher-level protocols and serves as the base layer in the emerging multi-protocol agent stack. It anchors interoperability by enabling consistent, reliable access to external tools and structured context.
- **Agent-to-Agent Protocol (A2A)**: Developed by Google, A2A supports secure, dynamic collaboration between heterogeneous agents. It relies on a structured JSON-based lifecycle model to describe tasks, agent capabilities, and shared artifacts, enabling peer-to-peer task outsourcing and decentralized workflows. Operating on a many-to-many communication model, A2A facilitates multimodal interaction and is particularly well-suited for distributed coordination across enterprise-scale systems. It builds upon the capabilities provided by MCP, using standardized tool and context access as a foundation for richer, collaborative execution between agents across varied technology stacks.

- **Agent Communication Protocol (ACP)**: ACP introduces a REST-native, performative messaging layer designed for local coordination between agents. It supports multipart messages and asynchronous streaming, providing a flexible communication interface for agents that are already integrated through MCP. Governed by the Linux Foundation, ACP is particularly useful for orchestrating multimodal interactions in scalable, message-rich environments. By layering REST-compliant messaging on top of MCP's context-sharing capabilities, ACP allows agents to communicate fluidly in structured and scalable workflows. It represents a critical middle layer in the protocol stack, sitting between MCP and higher-level collaboration protocols like A2A.
- **Agent Network Protocol (ANP)**: ANP is intended for cross-platform, internet-scale agent collaboration. It features a layered architecture that incorporates decentralized identity (via W3C DIDs), semantic web principles, and encrypted communication. ANP supports secure agent discovery, interaction, and coordination across open, decentralized environments, such as agent marketplaces or federated networks. Positioned as the top layer in the interoperability stack, ANP builds on MCP, ACP, and A2A to enable global-scale agent ecosystems. Its emphasis on decentralized identity and semantic interoperability allows agents to function independently across domains while maintaining trust and compatibility.

The interplay between these protocols suggests a progressive layering strategy for agent-based systems. MCP acts as the foundational layer focused on standardized access to tools and contextual data. ACP complements this by introducing robust message exchange infrastructure. A2A builds on both by enabling dynamic, task-centric peer interaction, while ANP extends interoperability to the open internet through decentralized identity and platform-agnostic semantics. This layered progression reflects an emerging model where context management (via MCP) is the base layer, followed by messaging (via ACP), collaboration (via A2A), and global interoperability (via ANP) [34].

This broader ecosystem of protocols signals a shift toward a modular, layered architecture for agent interoperability—similar in spirit to the protocol stacks that underpin the modern internet. While MCP anchors the system with reliable context and tool access, additional protocols like A2A, ACP, and ANP expand functionality to encompass communication, coordination, and global-scale agent networks. Rather than relying on a one-size-fits-all approach, the future of multi-agent systems appears to lie in a multi-protocol stack, where each layer addresses a different dimension of interoperability and enables agents to operate fluidly across increasingly complex and distributed environments.

## 5. Design Patterns in Practice: Architecting Inter-Agent Communication with MCP

The deployment of LLM-driven multi-agent systems is greatly enhanced by thoughtfully leveraging proven software design patterns, with the Model Context Protocol (MCP) playing a key role as the interoperability foundation that connects agents with tools, data, and one another.

### 5.1. Centralized Communication Architectures with MCP Mediation

**Description:** In centralized communication models, agents coordinate through a single control point or orchestrator. While this structure simplifies coordination in smaller-scale systems, it may introduce performance limitations as the number of agents grows or tasks become more complex [35][36].

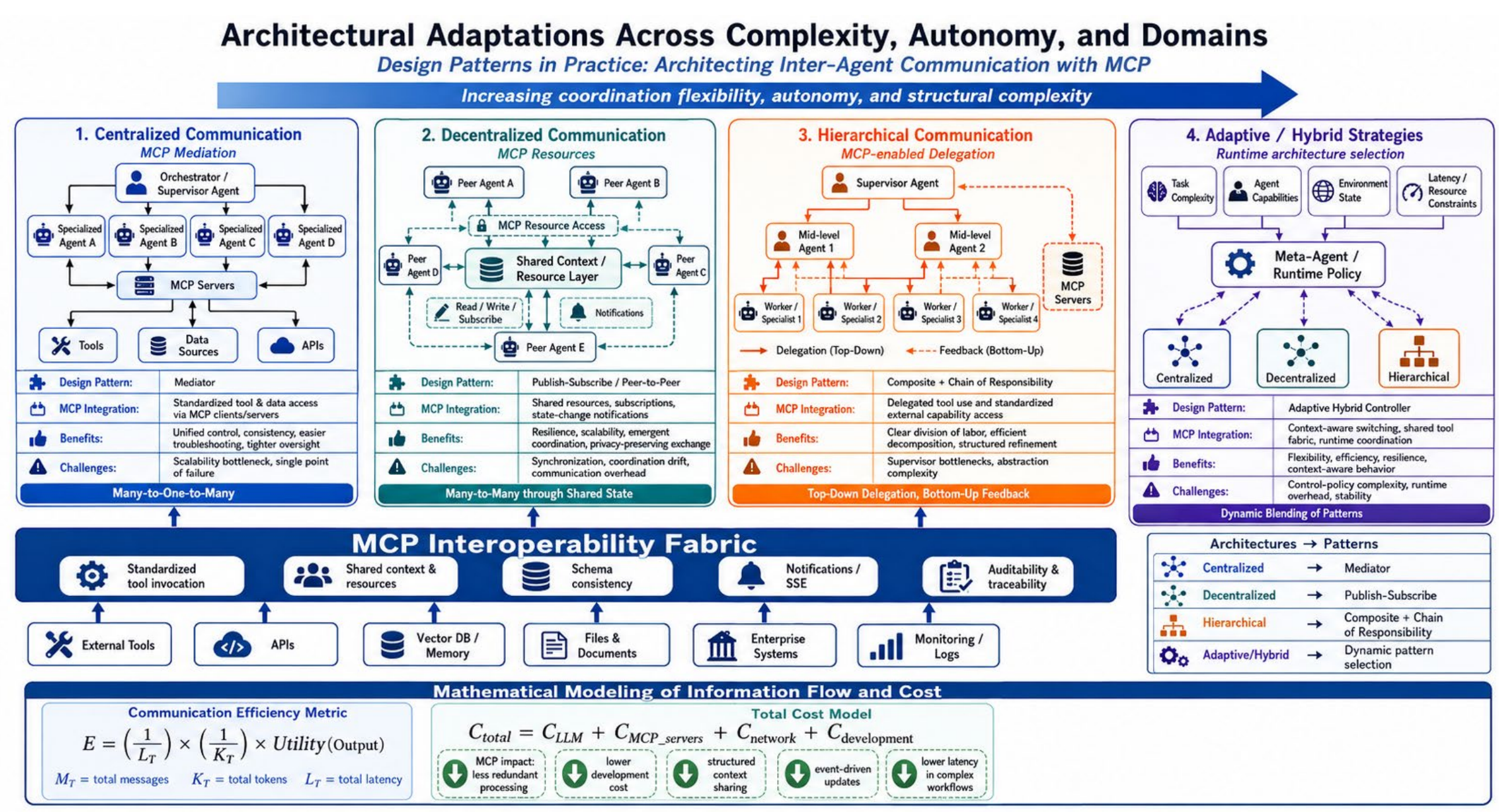

*Figure 3. Design patterns for inter-agent communication in LLM-based multi-agent systems, showing how centralized, decentralized, hierarchical, and adaptive architectures can all be grounded by MCP as a common interoperability fabric. The figure highlights communication flow, delegation structure, shared-state coordination, and a cost-effective view, illustrating why MCP enables a scalable, auditable, and production-ready agent ecosystem.*

**Design Pattern:** This model closely reflects the Mediator Pattern. A central decision-making agent—often an LLM—manages interactions by determining which specialized agent to engage next. These agents or their functions can be treated as callable tools, invoked based on the supervisor's planning logic.

**MCP Integration:** Within this framework, MCP servers serve as intermediaries that provide standardized access to tools and data across all agents. The central LLM-based orchestrator communicates with these MCP servers via clients, allowing it to issue tool calls and retrieve contextual data without handling tool-specific implementation details. This abstraction ensures uniform access to external resources and maintains system consistency [37].

**Benefits:** Centralized designs facilitate unified control, streamlined output consistency (often through centralized knowledge access), simplified troubleshooting, and tighter oversight of data and communication flows.

**Challenges:** However, reliance on a single orchestrator can create scalability constraints and pose risks associated with single points of failure.

**Conceptual Diagram: Centralized MCP-mediated Communication Flow**

Code snippet

```
graph TD
    subgraph LLM-MAS
        O
        A1(Specialized LLM Agent 1)
        A2(Specialized LLM Agent 2)
        A3(Specialized LLM Agent 3)

        O --- A1
        O --- A2
        O --- A3
    end

    subgraph External Environment
        MCPS
        ETDS(External Tools/Data Sources)

        MCPS --- ETDS
    end

    A1 -- MCP Client --> MCPS
    A2 -- MCP Client --> MCPS
    A3 -- MCP Client --> MCPS
    O -- MCP Client --> MCPS

    style O fill:#f9f,stroke:#333,stroke-width:2px
    style A1 fill:#bbf,stroke:#333,stroke-width:2px
    style A2 fill:#bbf,stroke:#333,stroke-width:2px
    style A3 fill:#bbf,stroke:#333,stroke-width:2px
    style MCPS fill:#ccf,stroke:#333,stroke-width:2px
    style ETDS fill:#cfc,stroke:#333,stroke-width:2px
```

*Description:* This diagram illustrates a centralized communication architecture. A central "Orchestrator/Supervisor Agent" (LLM) manages and directs multiple "Specialized LLM Agents." The Orchestrator communicates directly with these Specialized Agents. All Specialized Agents, and often the Orchestrator itself, interact with a pool of "MCP Servers," which in turn provide standardized access to "External Tools/Data Sources." The connections between agents and the orchestrator represent direct message passing or command delegation, while connections to MCP Servers represent MCP calls, with the MCP Servers acting as a central broker for external interactions.

## 5.2. Decentralized Communication Architectures Leveraging MCP Resources

**Description:** Decentralized architectures [38] emphasize peer-to-peer communication, distributing the communication load and eliminating single points of failure inherent in centralized models. In such systems, agents can specialize dynamically and route tasks without relying on rigid, predefined workflows.

**Design Pattern:** This approach is well-suited for the **Publish-Subscribe Pattern**. Agents publish messages to specific topics or channels, and other agents that have subscribed to those topics consume the relevant messages. While this fosters diverse ideas and emergent behavior, it can introduce synchronization challenges.

**MCP Integration:** MCP's resource capability is instrumental in enabling a decentralized Publish-Subscribe model. This feature allows agents to share various types of context—including files, application state, and agent memory—and, crucially, to subscribe to notifications when these shared resources change. This mechanism facilitates a decentralized coordination model where agents react to changes in shared state rather than relying on direct, explicit messages, thereby promoting asynchronous and loosely coupled interactions.

**Benefits:** Decentralized architectures offer greater resilience, enhanced scalability, the potential for emergent collective intelligence, and improved privacy preservation through minimal direct data exchange. **Challenges:** Key challenges include maintaining coordinated behavior across a distributed network, managing synchronization issues, and potentially higher overall communication overhead if not carefully managed [39].

**Conceptual Diagram: Decentralized MCP-mediated Communication Flow**

Code snippet

```
graph TD
    subgraph LLM-MAS
        A1(Specialized LLM Agent 1)
        A2(Specialized LLM Agent 2)
        A3(Specialized LLM Agent 3)

        A1 <--- Shared_Context_Layer ---> A2
        A2 <--- Shared_Context_Layer ---> A3
        A3 <--- Shared_Context_Layer ---> A1
    end

    subgraph External Environment
        MCPS(MCP Servers)
        ETDS(External Tools/Data Sources)

        MCPS --- ETDS
    end

    Shared_Context_Layer -- MCP Calls --> MCPS

    style A1 fill:#bbf,stroke:#333,stroke-width:2px
    style A2 fill:#bbf,stroke:#333,stroke-width:2px
    style A3 fill:#bbf,stroke:#333,stroke-width:2px
    style Shared_Context_Layer fill:#fcf,stroke:#333,stroke-width:2px
    style MCPS fill:#ccf,stroke:#333,stroke-width:2px
    style ETDS fill:#cfc,stroke:#333,stroke-width:2px
```

*Description:* This diagram illustrates a decentralized communication model in which several "Specialized LLM Agents" are interconnected and coordinate their actions through a common "Shared Context or Resource Layer." This shared layer might consist of a vector database, distributed storage, or other forms of persistent memory. Access to this layer is mediated by "MCP Servers," which offer a uniform interface for interaction. Agents communicate indirectly by using MCP to read from, write to, and subscribe to updates within this shared space. As agents modify resources, those changes become visible to others in real time, enabling coordination through a dynamically shared state rather than direct messaging..

## 5.3. Hierarchical Communication Architectures and MCP-enabled Delegation

**Conceptual Diagram: Hierarchical MCP-mediated Communication Flow**

Code snippet

```
graph TD
    subgraph LLM-MAS
        S(Supervisor Agent)
        M1(Mid-Level Agent 1)
        M2(Mid-Level Agent 2)
        W1(Specialized Worker Agent 1)
        W2(Specialized Worker Agent 2)
        W3(Specialized Worker Agent 3)
        W4(Specialized Worker Agent 4)

        S --- M1
        S --- M2
        M1 --- W1
        M1 --- W2
        M2 --- W3
        M2 --- W4
    end

    subgraph External Environment
        MCPS(MCP Servers)
        ETDS(External Tools/Data Sources)

        MCPS --- ETDS
    end

    W1 -- MCP Client --> MCPS
    W2 -- MCP Client --> MCPS
    W3 -- MCP Client --> MCPS
    W4 -- MCP Client --> MCPS
    M1 -- MCP Client --> MCPS
    M2 -- MCP Client --> MCPS
    S -- MCP Client --> MCPS

    style S fill:#f9f,stroke:#333,stroke-width:2px
    style M1 fill:#bbf,stroke:#333,stroke-width:2px
    style M2 fill:#bbf,stroke:#333,stroke-width:2px
    style W1 fill:#ccf,stroke:#333,stroke-width:2px
    style W2 fill:#ccf,stroke:#333,stroke-width:2px
    style W3 fill:#ccf,stroke:#333,stroke-width:2px
    style W4 fill:#ccf,stroke:#333,stroke-width:2px
    style MCPS fill:#cfc,stroke:#333,stroke-width:2px
    style ETDS fill:#fcf,stroke:#333,stroke-width:2px
```

**Description:** Hierarchical architectures organize agents into a tree-like structure, with higher-level agents overseeing broader objectives and delegating tasks to lower-level agents, or even a "supervisor of supervisors". This structure facilitates the division of labor among specialized agents and ensures their activities are synchronized to achieve overarching objectives [40].

**Design Pattern:** This pattern leverages elements of the **Composite Pattern** (for grouping agents into logical hierarchies) and the **Chain of Responsibility Pattern** (for sequential task delegation). Frameworks like "Talk Structurally, Act Hierarchically" (TalkHier) specifically introduce structured communication protocols and hierarchical refinement mechanisms to manage complexity.

**MCP Integration:** Higher-level agents can delegate specific sub-tasks or specialized data access requests to lower-level agents or directly to external tools via MCP. For instance, a manager agent might use an MCP client to invoke a tool (which could represent another agent's capability exposed as a tool) on an MCP server. This enables fine-grained control and efficient delegation of complex operations, allowing agents at different levels to leverage external capabilities in a standardized manner.

**Benefits:** Hierarchical systems offer streamlined decision-making, clear division of labor, efficient task decomposition, and improved refinement of outputs through structured feedback loops.
**Challenges:** Potential bottlenecks can arise at supervisor nodes, and managing multiple levels of abstraction can introduce architectural complexity.

**Conceptual Diagram: Hierarchical MCP-mediated Communication Flow**

*Description:* This diagram represents a hierarchical communication structure organized across multiple levels. At the highest level, a "Supervisor Agent" manages the overall workflow by assigning tasks to "Mid-Level Agents," who then distribute subtasks to "Specialized Worker Agents." The primary communication pattern follows a top-down and bottom-up flow. External integration is handled through "MCP Servers," which serve as standardized interfaces accessible to agents across the hierarchy. Higher-level agents may delegate external tool interactions to lower-level agents, allowing consistent and structured access to external systems throughout the entire architecture.

## 5.4. Adaptive and Hybrid Communication Strategies

In practice, the most effective communication strategy for LLM-based multi-agent systems (LLM-MAS) is seldom defined by a single, static architectural pattern. Instead, robust systems often integrate features from centralized, decentralized, and hierarchical models to accommodate varying levels of complexity and diverse operational demands. For example, a system may follow a mostly linear execution path but incorporate dynamic tool invocations at specific points to handle unpredictable tasks [9].

The key principle is to design communication strategies that align with system goals and current conditions. This often involves dynamically selecting or blending patterns based on factors such as task complexity, agent capabilities, or the state of the environment. Guidance from practitioners, such as Anthropic's suggestion to only increase architectural complexity when necessary, reinforces the idea that rigid, one-size-fits-all communication models are suboptimal. Instead, hybrid approaches that balance simplicity with flexibility tend to offer greater efficiency and resilience.

Moreover, the emergence of adaptive communication protocols highlights a shift toward systems that can modify their own coordination structures in real time. Rather than locking in communication strategies during system design, agents—or a dedicated meta-agent—can evaluate current conditions, such as changes in network layout, resource availability, or communication delays, and adapt the interaction model accordingly. This repositions communication architecture as a dynamic, runtime decision-making process rather than a fixed blueprint, enabling more intelligent, context-aware coordination across the agent network.

## 5.5. Mathematical Modeling of Inter-Agent Information Flow and Cost Optimization

To further optimize LLM-MAS, mathematical modeling can be applied to inter-agent information flow and cost.

**Communication Efficiency Metric**

Communication efficiency in an LLM-based multi-agent system can be quantitatively modeled by factoring in message volume, token usage, and response time.

Let $MT$ be the total number of messages exchanged in a system to complete a task.

Let $KT$ be the total number of tokens processed across all LLM calls.

Let $LT$ be the total latency for task completion.

An objective function for communication efficiency could be formulated as:

$$E=(1/LT)\times(1/KT)\times\text{Utility}(\text{Output})$$

The aim is to maximize this efficiency metric by producing high-utility outputs while keeping both the token count and latency as low as possible. This formulation encourages designs that are not only accurate and effective but also computationally and temporally efficient.

**MCP's Impact on Cost/Latency**

The Model Context Protocol (MCP) plays a central role in improving both cost-efficiency and latency in LLM-based multi-agent systems (LLM-MAS). By offering a unified and reusable interface for tool and context integration, MCP significantly lowers the need for repetitive development and minimizes ongoing maintenance demands. This translates into reduced initial engineering effort and lower long-term operational expenditures. While MCP introduces an intermediary layer, its reliance on lightweight JSON-RPC over HTTP and Server-Sent Events (SSE) often results in lower latency compared to ad hoc integration methods, particularly in complex system architectures.

The total cost of an MCP-enabled system (Ctotal) can be expressed as:

$$C_{total}=C_{LLM}+C_{MCP_servers}+C_{network}+C_{development}$$

Where:

- $C_{LLM}=\sum_i(tokens_{in,i}\times P_{in}+tokens_{out,i}\times P_{out})$ represents the cost of LLM inference, with $tokens_{in,i}$ and $tokens_{out,i}$ being input and output tokens for LLM call i, and $P_{in}, P_{out}$ being their respective costs per token.
- $C_{MCP_servers}$ represents the operational cost of running MCP servers.
- $C_{network}$ accounts for network data transfer costs.
- $C_{development}$ is the development cost, which is significantly reduced by MCP's standardization.

In multi-agent systems, where frequent inter-agent communication and LLM calls are expected, these costs can rise quickly. Token-based pricing makes LLM inference particularly sensitive to message volume and verbosity. MCP helps contain these expenses by minimizing redundant processing through standardized tool invocation, structured context sharing, and event-driven updates.

Mechanisms such as SSE for low-latency streaming and MCP's resource notification system further reduce unnecessary message duplication and network overhead. As a result, MCP not only simplifies integration and promotes modularity but also plays a vital role in the economic and performance scalability of multi-agent systems. Its efficiency-oriented design makes it a key enabler for transitioning LLM-MAS from experimental setups to cost-effective, production-grade deployments.

# 6. Architectural Adaptations Across Complexity, Autonomy, and Domains

## 6.1. Scaling Communication Patterns with Increasing Agent Complexity and Autonomy

LLM-based agent systems span a spectrum of complexity and autonomy—from simple, rule-based workflows to highly collaborative multi-agent ecosystems. As systems increase in sophistication, the communication patterns evolve, and the role of the Model Context Protocol (MCP) becomes more critical and multi-dimensional. MCP transitions from a lightweight integration layer into a foundational communication and interoperability infrastructure.

- **Low Complexity / Autonomy (Deterministic Chains / Workflows)**
    - **Description**: These systems follow fixed, sequential steps where the flow of execution is predefined. LLMs act more like orchestrators, invoking external tools in a linear fashion with minimal decision-making or dynamic adaptation.
    - **Communication**: Typically involves basic, stateless request-response interactions. There is little to no inter-agent communication, as workflows often consist of single-agent pipelines or simple Retrieval-Augmented Generation (RAG) chains.
    - **MCP Role**: At this level, MCP provides reliable, standardized access to external tools and data sources, ensuring consistent context retrieval for each step. It helps streamline integration and reduces redundant code for connecting to APIs or services. While its role is minimal, it offers tangible benefits in simplifying development and increasing reliability.
- **Medium Complexity / Autonomy (Single Agent with Dynamic Decisions)**
    - **Description**: Agents at this level begin to exhibit adaptive behavior, dynamically determining which tools to invoke and how to proceed based on intermediate outputs. Iterative reasoning, memory updates, and self-reflection are common [10].
    - **Communication**: Focused on internal agent loops—often involving "self-talk" (e.g., Chain-of-Thought), dynamic memory updates, and reactive tool calls depending on the agent's evolving understanding of its task.
    - **MCP Role**: MCP becomes essential for enabling dynamic and flexible tool interactions. It manages context updates across steps and provides consistent access to real-world information such as execution outputs, data queries, or code results. In doing so, it supports more autonomous decision-making and task refinement, helping the agent maintain a reliable view of its environment throughout execution.
- **High Complexity / Autonomy (Multi-Agent Architectures)**
    - **Description**: These systems consist of multiple specialized agents working collaboratively to solve more complex or distributed problems. Roles may be hierarchically organized or horizontally distributed, and require coordination, synchronization, and conflict resolution.
    - **Communication**: Involves rich inter-agent interaction using patterns like Mediator, Observer, Publish-Subscribe, Broker, Hierarchical, or Network-based topologies. Messaging can be asynchronous, stateful, and context-dependent, with agents needing shared understanding of goals and system state.
    - **MCP Role**: At this level, MCP functions as the interoperability backbone for the entire system. It enables standardized context sharing, event-driven updates, and tool invocation across agents built on potentially different frameworks or stacks. By abstracting the complexity of integration and enforcing consistent interfaces, MCP allows agents to operate as cohesive, coordinated entities. It supports advanced use cases like distributed task allocation, shared memory access, and reactive coordination through resource notifications.

As the autonomy and complexity of agent systems increase, the demands on their communication infrastructure also rise. MCP scales accordingly—from offering convenience and integration support in basic workflows to becoming the core layer that enables robust, scalable, and adaptive coordination in complex multi-agent environments. In simpler applications, MCP reduces boilerplate integration work. In dynamic, single-agent systems, it becomes the source of environmental grounding and dynamic execution. And in multi-agent ecosystems, it plays a pivotal role in enabling shared context, secure communication, and emergent behavior—making it indispensable for the next generation of agentic AI systems.

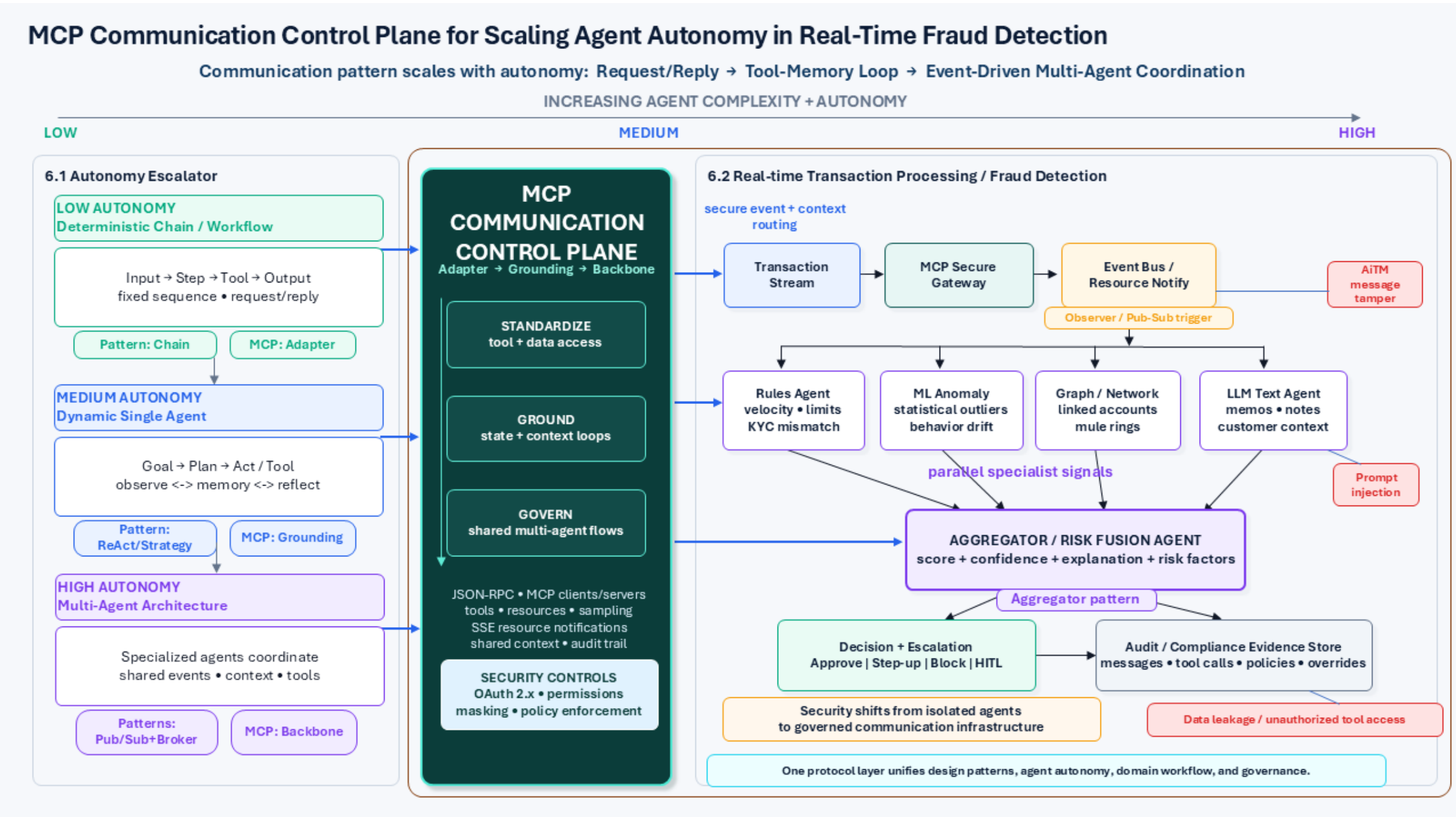

*Figure 4. MCP is the communication control plane that scales from adapter to grounding layer to governed multi-agent backbone; in real-time transaction processing, it routes events, coordinates fraud specialists, fuses risk through an Aggregator, and centralizes security/audit controls.*

## 6.2. Case Study: Real-time Transaction Processing Systems

**Domain Context**:

Real-time transaction processing—especially within financial systems—requires extremely high levels of precision, low-latency responses, strong security guarantees, and adherence to regulatory frameworks. LLM-based agents are increasingly used in these settings to support tasks such as analyzing transactional behavior, identifying potential fraud, and automating operational workflows.

**Key Challenges**:

Systems operating in this domain must handle sensitive financial data securely, provide agents with low-latency access to continuously updating transaction streams, and meet strict compliance requirements. Additionally, there is a need to avoid complex and fragile system architectures that can emerge from unstructured, ad hoc integrations—commonly referred to as “spaghetti code.”

**Design Patterns and MCP Application**:

- **Fraud Detection (Aggregator Pattern)**:
  In fraud detection systems, it’s common to deploy several specialized agents—each focused on different detection strategies, such as rules-based logic, statistical modeling, or network behavior analysis. These agents evaluate transactions independently and forward their outputs to an aggregator, which combines their inputs into a comprehensive fraud risk score. LLMs can support this pipeline by extracting nuanced insights from unstructured data, such as transaction memos or customer notes, transforming them into structured information aligned with an ontology, and contributing them to the broader analysis performed by other agents [41].

- **MCP Role**:
  MCP plays a key role in enabling fraud detection agents to securely and efficiently access transactional databases and real-time streams. It provides standardized, governed interfaces for tool and data integration, reducing the need for custom connectors. In addition, MCP can be used to generate synthetic datasets for safe testing of detection algorithms, helping developers mitigate risks related to data sensitivity while simulating realistic transaction scenarios.

- **Communication Flow**:
  Communication among agents in this context often adopts the Observer or Publish-Subscribe design pattern, enabling real-time fraud detection agents to react instantly to new transactions. MCP supports this model by offering event-driven resource change notifications. Additionally, MCP servers act as secure intermediaries in a Mediator or Broker pattern, facilitating controlled access to financial infrastructure. This abstraction not only simplifies the integration of complex banking systems but also ensures that all interactions remain secure, auditable, and policy-compliant.

**Security Implications**:
The rise of Agent-in-the-Middle (AiTM) attacks exposes a significant risk in the way messages are exchanged between agents in LLM-based multi-agent systems (LLM-MAS). This highlights the need for strong security mechanisms not just at the agent level but throughout the communication framework. MCP addresses these concerns through its support for OAuth 2.0/2.1 protocols for secure authentication and authorization, along with fine-grained access control and data masking features implemented at the server level. These capabilities are essential for maintaining the confidentiality, integrity, and regulated handling of sensitive data—particularly in domains such as finance [42].

In critical sectors like transaction processing, the combination of communication design patterns and MCP shifts the focus of system security from isolated agents to the infrastructure that connects them. Traditional approaches often concentrate on protecting each agent individually, but in LLM-MAS, the communication pathways themselves become a high-value target, as AiTM threats clearly illustrate. Even if agents are secure in isolation, the interception or alteration of messages between them can jeopardize the entire workflow. Patterns such as Mediator and Broker, when built on top of MCP, allow communication to be centralized or standardized, creating a single, auditable point of control.

This architectural shift enables security policies—such as token-based access, role-based permissions, and data redaction—to be enforced consistently across all interactions. By consolidating communication through MCP's structured and secure interfaces, systems can reduce the complexity and exposure of decentralized message exchanges. In doing so, MCP provides the infrastructure needed for safe, bidirectional communication that meets the stringent requirements of enterprise-grade financial systems [43].

### 6.3. Case Study: Investment Banking Applications

**Domain Context**:
Investment banking involves highly specialized decision-making that requires synthesizing large volumes of diverse data, operating under fast-changing market dynamics, and meeting strict risk and compliance standards. LLM-based agents are being developed to support tasks such as portfolio optimization, market trend analysis, advisory functions, and more complex workflows like mergers and acquisitions (M&A) [51][52].

**Key Challenges**:
The financial domain presents unique obstacles, including the need to process and interpret large-scale, multimodal data—such as text-based news, financial statements, and audio from earnings calls. Agents must also handle long-context reasoning, overcome the inherent opacity of traditional deep learning models, and operate within rigid regulatory frameworks requiring transparency and traceability.

**Design Patterns and MCP Application**:

- **FINCON Framework (Hierarchical Pattern)**:
  The FINCON [50] architecture reflects a hierarchical multi-agent setup modeled after the structure of investment firms. It organizes specialized agents—such as those focusing on sentiment analysis, technical indicators, or fundamentals—under a central decision-making agent that aggregates and evaluates their insights. The system also includes layered risk-control mechanisms designed to monitor and regulate exposure in real time.

- **Communication Flow**:
  FINCON minimizes unnecessary peer-to-peer chatter by channeling communication through structured hierarchies. It incorporates a selective signaling method—sometimes referred to as "verbal reinforcement"—to update only the agents impacted by new investment insights, thereby improving bandwidth and reasoning efficiency.

- **Portfolio Management (Parallel / Aggregator Patterns)**:
  In portfolio optimization scenarios, multiple agents may concurrently evaluate different types of risk (e.g., market volatility, sector correlation, credit exposure) or assess distinct asset categories. Their outputs are collected and synthesized by an aggregator agent to form a comprehensive view of portfolio health or risk.

- **MCP Role**:
  MCP enables agents to securely and consistently access up-to-date financial data from a variety of sources, including streaming market feeds, regulatory filings, and internal financial systems. It allows agents to subscribe to relevant updates and changes through resource notification mechanisms and ensures consistent data formatting across all interactions. MCP's context-sharing primitives ensure that agents remain synchronized with real-time market conditions—essential for high-frequency decision-making environments.

- **M&A (Complex Workflows)**:
  In M&A scenarios, a distributed team of agents could be deployed to handle various aspects of the process, such as financial due diligence, legal compliance, strategic alignment, and scenario modeling. These agents would coordinate using structured communication protocols and shared context layers, all supported through MCP's interoperable infrastructure. This ensures each specialized agent can contribute its insights to a larger decision-making process in a secure and transparent way.

LLM-MAS architectures in finance—especially those enhanced with MCP—are increasingly focused on delivering not just accurate decisions, but also explainable and auditable reasoning. Traditional AI models often fall short in clarity, making their outputs difficult to justify in regulated financial environments. Multi-agent systems like FINCON address this by structuring internal communication and reasoning flows around human-interpretable, evidence-based outputs. The manager-analyst hierarchy ensures that complex insights are broken down and expressed in clear language, while decisions are supported by traceable data. MCP plays a pivotal role in this process by providing structured access to the underlying data sources and tools, allowing agents to explicitly reference the foundations of their conclusions. Together, these mechanisms are helping shift financial AI systems from opaque black boxes to transparent, trustworthy, and regulation-ready platforms.

### 6.4. Financial Services Use Cases and Corresponding LLM-MAS Design Patterns

- **Fraud Detection**
  Fraud detection in financial systems leverages design patterns such as Aggregator, Observer/Publish-Subscribe, and Mediator/Broker. These patterns enable specialized detection agents—each using different methods like rule-based checks, machine learning, or behavioral analysis—to operate in parallel, share findings, and contribute to a consolidated fraud score. Communication is driven by real-time transaction events, which trigger agent responses via event streams, while structured inputs (such as "intuitive" language-based hunches) are passed through a central mediator or broker. MCP plays a key role by providing secure, real-time access to transactional databases and internal financial records, while also supporting the generation of synthetic data for testing in development environments. This architecture improves fraud detection speed, lowers false positive rates, and enables more proactive risk prevention. However, challenges persist around securing inter-agent communication against Agent-in-the-Middle (AiTM) attacks, handling highly sensitive financial data, and ensuring adherence to regulatory standards.

- **Portfolio Management**
  Portfolio management systems adopt hierarchical structures with manager-analyst models, along with parallel processing and aggregator patterns. Specialized agents may simultaneously assess market trends, asset risk, or credit exposure, feeding their insights into a centralized manager agent that generates portfolio-level strategies. Communication is streamlined through structured hierarchies and "verbal reinforcement"—a mechanism for propagating key investment updates to relevant agents without redundant messaging. MCP facilitates these workflows by offering uniform, real-time access to external data sources such as market feeds, financial disclosures, and internal records. It also supports context sharing to keep agents aligned with current portfolio conditions. The result is more efficient decision-making, deeper risk profiling, and data-backed investment strategies. Nevertheless, the dynamic nature of financial markets, combined with the complexity of synthesizing data from diverse sources, creates significant challenges in maintaining decision explainability and system robustness.

- **Financial Advisory**
  In financial advisory applications, LLM agents are designed to support hybrid and adaptive reasoning, often combining procedural strategies with contextual adjustments. These systems rely on real-time communication channels, context-aware dialogue, and human-in-the-loop mechanisms for resolving complex or personalized financial queries. MCP enables this use case by ensuring secure and standardized access to personal financial data, including client histories, investment goals, and market conditions. It allows advisory agents to generate personalized recommendations grounded in real-time data and secure client context. Benefits include more tailored advice, improved client trust, and adaptive learning paths for different financial profiles. The main challenges lie in maintaining conversational context over extended interactions, managing sensitive personally identifiable information (PII), and addressing the ethical implications of automated financial advice.

- **Mergers & Acquisitions (M&A) Due Diligence**
Though often implicit, M&A due diligence benefits from complex agent orchestration using composite, sequential, and workflow-based design patterns. Specialized agents focus on specific areas such as legal analysis, financial modeling, or strategic evaluation, each contributing to a shared understanding of the acquisition target. Communication follows a structured model involving document sharing, common knowledge repositories, and negotiation protocols. MCP supports this ecosystem by enabling secure and standardized access to data rooms, legal filings, and financial disclosures. It ensures that agents can interact with sensitive content without compromising security or consistency. This setup enhances the speed and accuracy of due diligence processes, enabling more comprehensive risk assessments. Key difficulties include handling unstructured documents at scale, integrating knowledge from various professional domains, and satisfying legal and regulatory expectations throughout the deal lifecycle.

## 7. Challenges, Security, and Future Research Directions

### 7.1. Addressing Scalability, Reliability, and Security in MCP-enabled MAS

Deploying LLM-based multi-agent systems (LLM-MAS) in real-world environments requires careful attention to **scalability**, **reliability**, and **security** to ensure operational robustness and trustworthiness.

- **Scalability**:
Although multi-agent systems are naturally modular and well-suited for scaling, the communication burden can grow rapidly as the number of agents increases, potentially leading to substantial coordination overhead. MCP helps mitigate this issue by providing standardized interfaces for tool and data integration, which reduces the need for custom-built connectors. This modular, plug-and-play approach streamlines the process of expanding systems and managing more agents without exponentially increasing system complexity.

- **Reliability**:
LLMs can sometimes produce inconsistent outputs or demonstrate sensitivity to prompt variations, resulting in unreliable behavior. In a multi-agent context, systems can address these issues using strategies such as internal feedback loops, self-evaluation, and agent cross-checking. MCP contributes to improved reliability by enforcing structured, consistent communication flows and context management. By ensuring that all agents access a unified, well-defined information space, MCP reduces the chances of miscommunication or hallucinated responses.

- **Security**:
Protecting sensitive data is critical, particularly in enterprise or regulated settings. MCP incorporates security as a core design feature through mechanisms like explicit user consent, defined permission boundaries, granular access control, and visibility into tool usage. However, threats such as Agent-in-the-Middle (AiTM) attacks still present potential risks—especially in systems that rely on shared context across agents. These challenges require a robust communication infrastructure that actively safeguards against unauthorized access or data leakage.

The growing need for LLMs to operate on enriched context introduces a trade-off between **contextual richness** and **security exposure**. As LLM performance improves with more available information, so too does the risk of sensitive data being unintentionally exposed. To manage this balance, MCP embeds security controls directly into its architecture. This includes OAuth-based authentication, fine-grained access permissions, and host-managed communication oversight. Rather than treating security as a post-deployment add-on, MCP follows a "secure-by-design" philosophy—integrating protective measures into every layer of the communication process. This foundational approach is essential for safely deploying context-aware agents in high-stakes environments where privacy, auditability, and compliance are non-negotiable.

### 7.2. Ethical Considerations and Human-in-the-Loop Integration

The implementation of LLM-based multi-agent systems (LLM-MAS) brings forth important ethical considerations, particularly in areas such as transparency, fairness, and responsibility. Without proper safeguards, LLMs may unintentionally propagate biases or inaccuracies embedded in their training data, potentially leading to flawed or unfair outcomes.

To mitigate these risks, incorporating **Human-in-the-Loop (HITL)** mechanisms is essential. Human oversight allows for review, intervention, or direct control over agent decisions—especially critical in domains where errors carry serious implications, such as healthcare, finance, or legal advisory. The Model Context Protocol (MCP) supports this mode of interaction by allowing MCP servers to pause agent execution and request additional user input, confirmation, or explicit consent. This capability ensures that automated decisions can be guided or overridden by human judgment when needed.

By embedding human oversight into agent workflows through MCP, the paradigm shifts from pure autonomy to **collaborative intelligence**—where humans and AI agents operate as integrated partners. Although LLM agents are designed for independent reasoning, their current limitations—including hallucinations, lack of ethical reasoning, and occasional unreliability—highlight the need for structured human involvement. HITL and related patterns like "human-on-the-loop" introduce intentional checkpoints for validation, context clarification, or ethical scrutiny.

MCP enables this by supporting interaction models that go beyond simple task execution. Features such as **elicitation** (the ability for servers to prompt for human feedback or input), fine-grained consent management, and user-directed control allow humans to stay actively involved in the loop. As a result, LLM-MAS evolve from standalone autonomous agents into systems that emphasize **shared decision-making**. This vision redefines communication protocols not just as pathways for inter-agent collaboration, but as bridges for seamless, secure, and transparent human-AI cooperation—ensuring that accountability and trust remain central in increasingly intelligent and complex systems [53].

### 7.3. Open Research Questions and Emerging Trends

The landscape of LLM-driven agentic AI and multi-agent systems is advancing rapidly, giving rise to a range of unresolved research challenges and emerging directions:

- **Defining Formal Semantics for Agent Communication**
  Traditional Agent Communication Languages (ACLs), such as FIPA-ACL, were built on strict formal semantics to ensure clarity and consistency. In contrast, LLM-based agents often communicate using natural language, which, while flexible and expressive, introduces ambiguity. A key research goal is to bridge the gap between natural language communication and the level of precision required for reliable, machine-to-machine

interaction—especially in systems where misinterpretation could lead to cascading errors [54].

- **Incorporating Multi-Modal Communication**
  A growing research frontier involves expanding agent capabilities to interpret and exchange information across multiple modalities—such as visual, auditory, and textual data. Enabling agents to engage through combinations of text, images, or speech can significantly enhance their ability to perceive context, interpret complex scenarios, and make better-informed decisions.

- **Real-Time Role Adaptation and Self-Organizing Architectures**
  Another important research focus is enabling agents to reorganize themselves dynamically in response to evolving goals or environmental shifts. This includes forming task-specific teams, redistributing responsibilities, and adjusting communication flows on the fly. Such adaptability moves beyond rigid system configurations, allowing agents to operate in more fluid, open-ended environments [55].

- **Benchmarking and Evaluation**:
  There is an urgent need for evaluation frameworks that go beyond measuring the performance of individual agents and instead assess how effectively multiple agents work together. Such benchmarks should capture aspects like coordination efficiency, collective reasoning, task distribution, and emergent behaviors in complex environments [56].

- **Long-term Learning and Adaptation**:
  A major open challenge is enabling multi-agent systems to continuously evolve—learning new strategies, refining communication protocols, and adapting their behavior as environments and tasks change over time. This includes the ability to retain useful knowledge, respond to feedback, and improve coordination strategies in non-static, real-world scenarios.

- **Decentralized Agent Marketplaces**:
  Emerging efforts around protocols like the Agent Network Protocol (ANP) suggest a future where agents can be discovered, composed, and deployed across open networks. These systems envision secure, decentralized environments where independently developed agents can collaborate, negotiate roles, and transact capabilities without centralized control.

Looking ahead, the evolution of LLM-based multi-agent systems points toward architectures that are **inherently adaptive and self-organizing**, far beyond the limitations of traditional software design models. Today's systems typically rely on predefined roles and fixed communication flows—even in complex tasks. However, future systems are expected to dynamically select, modify, or even invent new coordination strategies in response to previously unseen situations. This progression implies the emergence of **meta-level AI mechanisms** that govern not just behavior, but the underlying design architecture itself.

Such developments could lead to the rise of **AI-generated design patterns**, where system configuration becomes a fluid, evolving process rather than a static engineering choice. This shift marks a fundamental transformation—from applying established software principles to building systems capable of autonomously redefining their own communication and coordination logic, bridging the gap between human-designed architectures and autonomous AI systems.

## 8. Conclusion

The accelerating adoption of Large Language Model (LLM)-driven agentic AI systems calls for a structured and thoughtful approach to inter-agent communication—an essential element that directly impacts the effectiveness of multi-agent coordination. This review has shown that classical software design patterns such as Mediator, Observer, Publish-Subscribe, and Broker continue to offer valuable architectural guidance. These patterns support critical features like decoupled interaction, asynchronous updates, and modular design, making them well-suited for building scalable, maintainable, and interoperable multi-agent systems.

Within this landscape, the Model Context Protocol (MCP) plays a foundational role as a unifying interoperability layer. By introducing standardized mechanisms for context sharing and tool invocation, MCP reduces integration complexity and supports the deployment of design patterns across diverse LLM-based agents. It addresses long-standing integration bottlenecks—such as the "N x M" problem of connecting multiple models to various tools—by enabling seamless and secure access to external systems. This structured communication layer empowers agents to function collaboratively across a wide range of system architectures and autonomy levels.

The real-world advantages of combining classical design patterns with MCP are particularly evident in complex, regulated domains. In real-time transaction environments, for example, architectural strategies like the Aggregator pattern—paired with MCP's secure data access capabilities—enhance fraud detection workflows while supporting compliance requirements. In investment banking, hierarchical multi-agent frameworks such as FINCON demonstrate how agent specialization, structured communication, and MCP-driven data access can enable sophisticated decision-making for tasks like market forecasting and portfolio management, all while contributing to greater model transparency and explainability.

Despite ongoing challenges related to scalability, consistency, and security—especially within the communication layer—progress is being made through the integration of human oversight mechanisms and the emergence of new interoperability protocols like A2A, ACP, and ANP. Together, these advancements are pushing LLM-MAS toward systems that are not only more robust and interpretable but also aligned with ethical and regulatory standards.

Looking ahead, the field is moving beyond static communication patterns toward architectures that are adaptive, context-aware, and self-organizing. These dynamic systems promise to evolve their communication strategies in real time, redefining how coordination is designed and executed in AI ecosystems. Ultimately, the transformative potential of LLM-based multi-agent systems will depend not solely on the intelligence of individual models, but on the intelligent structuring of their collaborative behaviors and interactions.

## References

[1] T. Zhang, "System architecture for agentic large language models," UC Berkeley Electronic Theses and Dissertations, 2024. [Online]. Available: https://escholarship.org/uc/item/5zt3h2r2

[2] SuperAnnotate, "LLM agents: The ultimate guide," Web blog post, 2025. [Online]. Available: https://www.superannotate.com/blog/llm-agents

[3] W. Talukdar, "Autonomous AI Agents: Leveraging LLMs for Adaptive Decision-Making in Real-World Applications,"� IEEE Computer Society, Community Voices, 2025.plicat

[4] G. D. A. E. Aquino, R. Gomes, and I. G. TornÃ, "From RAG to Multi-Agent Systems: A Survey of Modern Approaches in LLM Development," Preprints, 2025. [Online]. Available: https://doi.org/10.20944/preprints202502.0406.v1

[5] Yan et al., "Beyond Self-Talk: A Communication-Centric Survey of LLM-Based Multi-Agent Systems," arXiv preprint arXiv:2502.14321, 2025.

[6] LLM Agents, "Prompt Engineering Guide," 2025. [Online]. Available: https://www.promptingguide.ai/research/llm-agents

[7] k2view, "What are LLM Agents?," 2025. [Online]. Available: https://www.k2view.com/what-are-llm-agents/

[8] K.-T. Tran et al., "Multi-Agent Collaboration Mechanisms: A Survey of LLMs," arXiv preprint arXiv:2501.06322, 2025.

[9] Engineering at Anthropic, "Building effective agents," 2024. [Online]. Available: https://www.anthropic.com/engineering/building-effective-agents

[10] LangChain, ""How to think about agent frameworks,""� 2025. [Online]. Available: https://blog.langchain.dev/how-to-think-about-agent-frameworks/

[11] K.-T. Tran et al., "Multi-Agent Collaboration Mechanisms: A Survey of LLMs," arXiv preprint arXiv:2501.06322, 2025.

[12] Sandeep, "MCP – A Beginner's Guide." 2025. [Online]. Available: url:https://opencv.org/blog/model-context- protocol/

[13] k2view. "What is Model Context Protocol?. 2025. https://www.k2view.com/model-context- protocol/

[14] descope. What Is the Model Context Protocol (MCP) and How It Works. 2025. https://www.descope.com/learn/post/mcp

[15] LLM Agent Frameworks 2025: Guide & Comparison. 2025. url:https://www.chatbase.co/blog/llm-agent-framework-guide

[16] Agent system design patterns. databricks. 2025.

https://docs.databricks.com/aws/en/generative-ai/guide/agent-system-design-patterns

[17] Cem Dilmegani, Mert Palazoğlu. 4 Agentic AI Design Patterns & Real-World Examples. 2025. url:https://research.aimultiple.com/agentic-ai-design-patterns/

[18] Bingyu Yan, , Xiaoming Zhang, Litian Zhang, Lian Zhang, Ziyi Zhou, Dezhuang Miao, and Chaozhuo Li. "Beyond Self-Talk: A Communication-Centric Survey of LLM-Based Multi-Agent Systems." (2025).

[19] Mert Cemri, , Melissa Z. Pan, Shuyi Yang, Lakshya A. Agrawal, Bhavya Chopra, Rishabh Tiwari, Kurt Keutzer, Aditya Parameswaran, Dan Klein, Kannan Ramchandran, Matei Zaharia, Joseph E. Gonzalez, and Ion Stoica. "Why Do Multi-Agent LLM Systems Fail?." (2025).

[20] Multi-agent systems.LangGraph.2025.url:https://langchain-ai.github.io/langgraph/concepts/multi_agent/

[21] Deloitte. The cognitive leap How to reimagine work with AI agents. 2024. rl:https://www2.deloitte.com/content/dam/Deloitte/us/Documents/gen-ai-multi-agents-pov-2.pdf

[22] Zeng, Y., C., Brown, J., Raymond, M., Byari, R., Hotz, and M., Rounsevell. "Exploring the opportunities and challenges of using large language models torepresent institutional agency in land system modelling".Earth System Dynamics 16, no.2 (2025): 423–449.

[23] Guannan Liang, , and Qianqian Tong. "LLM-Powered AI Agent Systems and Their Applications in Industry." (2025).

[24] Lance Johnson. MCP broker pattern: Key concepts & applications in modern distributed systems. 2025. url:https://www.byteplus.com/en/topic/542191?title=mcp-broker-pattern-key-concepts-applications-in-modern-distributed-systems

[25] Zeeshan, Talha, et al. Large Language Model Based Multi-Agent System Augmented Complex Event Processing Pipeline for Internet of Multimedia Things. 2025. doi:10.48550/arXiv.2501.00906 (2025).

[26] Understanding the Model Context Protocol: Architecture. 2025. url:https://nebius.com/blog/posts/understanding-model-context-protocol-mcp-architecture

[27] modemcontextprotocol.Core architecture.2025.url:https://modelcontextprotocol.io/docs/concepts/architecture

[28] geeksforgeeks.Broker Pattern.2025.url:https://www.geeksforgeeks.org/broker-pattern/

[29] Edwin Lisowski.Model Context Protocol (MCP): Solution to AI Integration Bottlenecks.2025.url:https://addepto.com/blog/model-context-protocol-mcp-solution-to-ai-integration-bottlenecks/

[30] Nick Aldridge, Marc Brooker, and Swami Sivasubramanian. Open Protocols for Agent Interoperability Part 1: Inter-Agent Communication on MCP. 2025. AWS Open Source Blog. url:https://aws.amazon.com/blogs/opensource/open-protocols-for-agent-interoperability-part-1-inter-agent-communication-on-mcp/

[31] Yusen Zhang, Student Researcher, and Ruoxi Sun. Chain of Agents: Large language models collaborating on long-context tasks. 2025. Google Research. url:https://research.google/blog/chain-of-agents-large-language-models-collaborating-on-long-context-tasks/

[32] Sijie Guo. Open Standards for Real-Time AI Integration – A Look at MCP. 2025. url:https://streamnative.io/blog/open-standards-real-time-ai-mcp

[33] Abul Ehtesham, , Aditi Singh, Gaurav Kumar Gupta, and Saket Kumar. "A survey of agent interoperability protocols: Model Context Protocol (MCP), Agent Communication Protocol (ACP), Agent-to-Agent Protocol (A2A), and Agent Network Protocol (ANP)." (2025).

[34] Ofir Yakobi,Shir Sadon. Bringing Memory to AI: A Look at A2A and MCP-like Technologies Across Platforms. 2025. url:https://orca.security/resources/blog/bringing-memory-to-ai-mcp-a2a-agent-context-protocols/

[35] Spheron Network.Building Your First Hierarchical Multi-Agent System.2025.url:https://blog.spheron.network/building-your-first-hierarchical-multi-agent-system

[36] Stuti Dhruv, Pawan Pawar.How to Build Multi Agent AI System.2025.https://www.aalpha.net/blog/how-to-build-multi-agent-ai-system/

[37] Isabelle Nguyen,Tanay Soni.Understanding the Model Context Protocol (MCP).2025.url:https://www.deepset.ai/blog/understanding-the-model-context-protocol-mcp

[38] Yingxuan Yang, Huacan Chai, Shuai Shao, Yuanyi Song, Siyuan Qi, Renting Rui, & Weinan Zhang. (2025). AgentNet: Decentralized Evolutionary Coordination for LLM-based Multi-Agent Systems.

[39] Khanh-Tung Tran, Dung Dao, Minh-Duong Nguyen, Quoc-Viet Pham, Barry O'Sullivan, & Hoang D. Nguyen. (2025). Multi-Agent Collaboration Mechanisms: A Survey of LLMs.

[40] Zhao Wang, Sota Moriyama, Wei-Yao Wang, Briti Gangopadhyay, & Shingo Takamatsu. (2025). Talk Structurally, Act Hierarchically: A Collaborative Framework for LLM Multi-Agent Systems.

[41] Hasan Mehdi,Alfredo Castillo. Multi-Agent System Patterns in Financial Services: Architectures for Next-Generation AI Solutions. 2025. url:https://community.aws/content/2uDxjoo105xRO6Q7mfkogmOYTVp/multi-agent-system-patterns-in-financial-services-architectures-for-next-generation-ai-solutions

[42] Gonzalo Sanchez. What is Model Context Protocol and how to leverage it in the fintech industry?. 2025. url:https://prometeoapi.com/en/blog/model-context-protocol-fintech

[43] He, P., Lin, Y., Dong, S., Xu, H., Xing, Y., & Liu, H. (2025). Red-Teaming LLM Multi-Agent Systems via Communication Attacks. *Preprint*.

[44] Soumyendu Sarkar, Sahand Ghorbanpour, Ricardo Luna Gutierrez, Antonio Guillen-Perez, Vineet Gundecha, Ashwin Ramesh Babu, Avisek Naug. (2025). Review of LLM based Control. url:https://www.researchgate.net/publication/390334927_Review_of_LLM_based_Control

[45] Soumyendu Sarkar, Sahand Ghorbanpour, Ricardo Luna Gutierrez, Antonio Guillen-Perez, Vineet Gundecha, Ashwin Ramesh Babu, Avisek Naug. Review of LLM based Control. (2025). url: https://www.researchgate.net/publication/390334927_Review_of_LLM_based_Control

[46] Sahand Ghorbanpour, Ricardo Luna Gutierrez, Vineet Gundecha, Desik Rengarajan, Ashwin Ramesh Babu, Soumyendu Sarkar. LLM Enhanced Bayesian Optimization for Scientific Applications like Fusion. (2024). url: https://openreview.net/forum?id=VnxFb2VG4E

[47] Mousavi, S., Guti'errez, R.L., Rengarajan, D., Gundecha, V., Babu, A.R., Naug, A., Guillen-Perez, A., Sarkar, S. N-critics: Self-refinement of large language models with ensemble of critics. (2023). ArXiv abs/2310.18679

[48] Sajad Mousavi, Desik Rengarajan, Ashwin Ramesh Babu, Sahand Ghorbanpour, Vineet Gundecha, Avisek Naug, Soumyendu Sarkar. Informed Tree of Thought: Cost-efficient Problem Solving with Large Language Models. (2024). url: https://openreview.net/forum?id=EJNkaV27yB

[49] Sajad Mousavi, Desik Rengarajan, Ashwin Ramesh Babu, Vineet Gundecha, Antonio Guillen, Ricardo Luna Gutierrez, Avisek Naug, Sahand Ghorbanpour, Soumyendu Sarkar. Imitation Guided Automated Red Teaming. (2024). url: https://openreview.net/forum?id=STIC40U1es

[50] Yangyang Yu, Zhiyuan Yao, Haohang Li, Zhiyang Deng, Yupeng Cao, Zhi Chen, Jordan W. Suchow, Rong Liu, Zhenyu Cui, Zhaozhuo Xu, Denghui Zhang, Koduvayur Subbalakshmi, Guojun Xiong, Yueru He, Jimin Huang, Dong Li, & Qianqian Xie. (2024). FinCon: A Synthesized LLM Multi-Agent System with Conceptual Verbal Reinforcement for Enhanced Financial Decision Making.

[51] Yijia Xiao, , Edward Sun, Di Luo, and Wei Wang. "TradingAgents: Multi-Agents LLM Financial Trading Framework." (2025).
[52] Li, Haohang, Yupeng, Cao, Yangyang, Yu, Shashidhar Reddy, Javaji, Zhiyang, Deng, Yueru, He, Yuechen, Jiang, Zining, Zhu, Koduvayur, Subbalakshmi, Guojun, Xiong, Jimin, Huang, Lingfei, Qian, Xueqing, Peng, Qianqian, Xie, and Jordan, Suchow. "INVESTORBENCH: A Benchmark for Financial Decision-Making Tasks with LLM-based Agent".*preprint* (2024).

[53] Guannan Liang, , and Qianqian Tong. "LLM-Powered AI Agent Systems and Their Applications in Industry." (2025).
[54] Alexander De Ridder. Agent Communication Language Definition: Understanding Its Role in Multi-Agent Systems. 2025. url:https://smythos.com/ai-agents/ai-agent-development/agent-communication-language-definition/

[55] Yingxuan Yang, , Huacan Chai, Shuai Shao, Yuanyi Song, Siyuan Qi, Renting Rui, and Weinan Zhang. "AgentNet: Decentralized Evolutionary Coordination for LLM-based Multi-Agent Systems." (2025).
[56] Li, Haohang, Yupeng, Cao, Yangyang, Yu, Shashidhar Reddy, Javaji, Zhiyang, Deng, Yueru, He, Yuechen, Jiang, Zining, Zhu, Koduvayur, Subbalakshmi, Guojun, Xiong, Jimin, Huang, Lingfei, Qian, Xueqing, Peng, Qianqian, Xie, and Jordan, Suchow. "INVESTORBENCH: A Benchmark for Financial Decision-Making Tasks with LLM-based Agent".*preprint* (2024).